\documentclass[10pt,preprint,aps,prb,citeautoscript,amssymb,showpacs,superscriptaddress,floatfix]{revtex4-1}
\usepackage{graphicx}
\usepackage{dcolumn}
\usepackage{bm}
\usepackage[T1]{fontenc}
\usepackage{lmodern}
\usepackage{natbib}
\usepackage{makecell}
\usepackage{booktabs}
\usepackage{multirow}
\usepackage{diagbox}

\usepackage{amsmath,amssymb}
\usepackage{graphicx,textcomp}
\usepackage{subfigure}
\usepackage{color}
\usepackage[colorlinks=true,
            linkcolor=blue,
            urlcolor=blue,
            citecolor=blue]{hyperref}

\renewcommand {\vec}    [1]    {\ensuremath{\mathbf{#1}}}

\newcommand   {\mat}    [1]    {\ensuremath{\mathbf{\bar{\bar{#1}}}} }			
\newcommand	  {\set}    [1]    {\ensuremath{ \{ #1 \} }}

\newcommand*{\citen}[1]{%
  \begingroup
    \romannumeral-`\x 
    \setcitestyle{numbers}%
    \cite{#1}%
  \endgroup   
}

\begin{document}

\title{Finite temperature elastic constants of paramagnetic materials within the disordered local moment picture from \textit{ab initio} molecular dynamics calculations}

\author{E. Mozafari}
\email[Electronic mail:] {elhmo@ifm.liu.se}
\affiliation{Department of Physics, Chemistry and Biology, Link\"{o}ping University, SE-58183 Link\"{o}ping, Sweden}

\author{N. Shulumba}
\affiliation{Department of Physics, Chemistry and Biology, Link\"{o}ping University, SE-58183 Link\"{o}ping, Sweden}

\author{P. Steneteg}
\affiliation{Department of Physics, Chemistry and Biology, Link\"{o}ping University, SE-58183 Link\"{o}ping, Sweden}

\author{B. Alling}
\affiliation{Department of Physics, Chemistry and Biology, Link\"{o}ping University, SE-58183 Link\"{o}ping, Sweden}
\affiliation{Max-Planck-Institut f\"{u}r Eisenforschung GmbH, 40237 D\"{u}sseldorf, Germany}

\author{Igor A. Abrikosov}
\affiliation{Department of Physics, Chemistry and Biology, Link\"{o}ping University, SE-58183 Link\"{o}ping, Sweden}
\affiliation{Materials Modeling and Development Laboratory, NUST "MISIS", 119049 Moscow, Russia}
\affiliation{LACOMAS Laboratory, Tomsk State University, 634050 Tomsk, Russia}

\date{\today}

\begin{abstract} 
We present a theoretical scheme to calculate the elastic constants of magnetic materials in the high-temperature paramagnetic state. Our approach is based on a combination of disordered local moments picture and \textit{ab initio} molecular dynamics (DLM-MD). Moreover, we investigate a possibility to enhance the efficiency of the simulations of elastic properties using recently introduced method: symmetry imposed force constant temperature dependent effective potential (SIFC-TDEP). We have chosen cubic paramagnetic CrN as a model system. This is done due to its technological importance and its demonstrated strong coupling between magnetic and lattice degrees of freedom. We have studied the temperature dependent single-crystal and polycrystalline elastic constants of paramagentic CrN up to 1200 K. The obtained results at T= 300 K agree well with the experimental values of polycrystalline elastic constants as well as Poisson ratio at room temperature. We observe that the Young's modulus is strongly dependent on temperature, decreasing by $\sim$14\% from T=300 K to 1200 K. In addition we have studied the elastic anisotropy of CrN as a function of temperature and we observe that CrN becomes substantially more isotropic as the temperature increases. We demonstrate that the use of Birch law may lead to substantial errors for calculations of temperature induced changes of elastic moduli. The proposed methodology can be used for accurate predictions of mechanical properties of magnetic materials at temperatures above their magnetic order-disorder phase transition.    
\end{abstract}

\pacs{71.15.Pd, 65.40.-b, 62.20.de}

\maketitle

\section{Introduction}
Elastic properties, an important part of the mechanical response of a material, are among the major properties to be studied in theorteical simulations. For instance, Barannikova \textit{et al}.\cite{Barannikova2012} have recently shown that there is a significant correlation between the elastic and plastic processes that are involved simultaneously in deforming alloys. 
It is known that for magnetic materials the existance of local magnetic moments above the magnetic transition temperature, in the paramgentic state noticeably affects the elastic properties\cite{Rivadulla2009a,Alling2010e}.  
Thus, a possibility to predict elastic moduli of magnetic materials in their high temperature paramagnetic state as a function of temperature is highly requested. \\ 
The main approach to incorporate temperature in theoretical studies of elastic properties of magnetic materials have been through approximations made in \textit{ab initio} schemes by excluding the implicit effect of lattice vibrations\cite{Leonov2012}, or by including the thermal expansion effects, often using the experimental data\cite{Zhang2011,Ruban2012}. While lattice expansion is believed to be the most important contribution to the temperature dependence of  elastic moduli, it would be worthwhile to develop methods which enable us to directly investigate the full effect of the temperature. Unfortunately, a consistent description of a paramagnetic state of a magnetic material is a highly non-trivial theoretical task\cite{Abrikosov2014}. In particular, we are not aware of any theoretical calculation of elastic moduli, where lattice vibrations and magnetic disorder are included on the same footing. However, this could be achieved by the disordered local moments molecular dynamics (DLM-MD)\cite{Steneteg2012a,Abrikosov2014}. \\
In this article, we have used DLM-MD method to study the temperature-dependent elastic moduli of paramagnetic B1 CrN chosen as our model system to demonstrate the functionality of our method. CrN, is a very interesting system considering both its industrial applications in hard coatings and its physical properties. Many experimental and theoretical groups have studied CrN because of its wide range of applications\cite{Corliss1960a,Gall2002a,Mayrhofer2008,Rivadulla2009a,Wang2012a,
Alling2010d,Steneteg2012a,Zhou2014}. Apart from its valuable applications, CrN exhibits some fascinating fundamental physical properties. Just below room temperature, $T_N\sim 270-286$ $K$, CrN undergoes a phase transition from an orthorhombic antiferromagnetic (AFM) phase to a cubic B1 paramagnetic (PM) phase\cite{Corliss1960a,Rivadulla2009a,Wang2012a}. The phase transition in CrN is due to magnetic entropy and the structural change is related to magnetic stress. The volume of the unit cell reduces by $\sim 0.59\%$ when CrN transforms from cubic to orthorhombic\cite{Corliss1960a,Browne1970a}. Shulumba \textit{et al} have used a combination of DLM-MD and TDEP to study vibrational free energy and phase stability of CrN\cite{Shulumba2014}. In their study the magnetic and vibrational degrees of freedom are coupled through the forces from DLM-MD. They found the transition temperature of CrN to be around 380 K which is a closer value to the experimental value of 286 K\cite{Rivadulla2009a} than the result obtained neglecting the effect of lattice vibrations. \\       
Among all the research that has been done on CrN, only a few has studied its elastic properties\cite{Alling2010e,Alling2010d,Steneteg2012a,Wang2012a,Zhou2014,Wang2016}. Importantly, theoretical studies of the elastic properties of CrN, up to this date, have been carried out only at $T=0$ $K$\cite{Zhou2014}. 
An important explanation for this is that until recently there were no appropriate computational tools to treat simultaneously effects of lattice vibrations and magnetic disorder. Thus, application of the proposed technique to study temperature dependent elastic properties of CrN is justified. 
In addition, we investigate the possibility to enhance the efficiency of the simulations of elastic properties and present the results obtained from the recently developed method, symmetry imposed force constants temperature dependent effective potential (SIFC-TDEP)\cite{Shulumba2015}, combined with the DLM picture, by relating the phonon properties of the magnetic disorder system and elasticity. 
Moreover, we compare both theoretical schemes with the results from more conventional calculations based on the Birch law in which the effect of temperature on the elastic moduli is introduced through the thermal expansion and discuss the accuracy of all the schemes considered in this study.
 
    
\section{Methodology}
\label{sec.2}
\subsection{Elastic Properties} \label{2A}
To simulate the paramagnetic phase we have used the disordered local moments molecular dynamics (DLM-MD)\cite{Steneteg2012a}. In this method, the local moments are spatially disordered and the magnetic state of the system is modified periodically and rearranged randomly with a specific time step, spin flip time, during the course of MD simulation. Using the experimental volumes at specific temperatures, we then apply five different deformations to the lattice and perform separate DLM-MD calculations for each deformation. More details on the DLM-MD method are given in Sec.~\ref{2B}.  \\
The stress-strain relation in the Voigt notation is defined as\cite{Grimvall1999}
\begin{equation}
\label{eq.1}
\sigma_{i} = \sum_{j} C_{ij} \epsilon_{j}
\end{equation}
where $\sigma$ is the stress tensor and $C_{ij}$s are the elements of the elastic tensor space.
In a cubic system, due to the symmetry, the elastic tensor will only have three non-vanishing, unidentical elements $C_{11}$, $C_{12}$ and $C_{44}$. To obtain the elastic constants, we have used the following deformation matrix\cite{Steneteg2013}
\begin{equation}
\label{eq.2}
\epsilon_{\eta} = 
 \begin{pmatrix}
  1+\eta & \eta/2 & 0 \\
  \eta/2 & 1 & 0 \\
  0 & 0  & 1  \\ 
 \end{pmatrix}
\end{equation}
Inserting this matrix into eq.~\ref{eq.1}, the elastic constants are derived as
\begin{equation}
\frac{d\sigma_1(T)}{d\eta}=C_{11}(T) \label{eq.5}
\end{equation}
\begin{equation}
\frac{d\sigma_2(T)}{d\eta}=C_{12}(T) \label{eq.6}
\end{equation}
\begin{equation}
\frac{d\sigma_6(T)}{d\eta}=C_{44}(T) \label{eq.7}
\end{equation}
First we calculate the stress, $\sigma$, for a set of deformations, $\epsilon(\eta)$, with $\eta$ deviating just a few percent, in our calculations 1\%, from zero. Then, we obtain the numerical derivative of $\sigma$ and extract the elastic constants. For each temperature and the volume at that temperature, we calculate stresses, $\sigma$, for a set of molecular dynamics time steps, $N_t=5000$. Thus, for each $\eta$, we will have $N_t$ number of stresses. The derivative is numerically calculated by fitting a line to these points using the least square method. \\
In principle, single crystal samples are often not available, thus the measurements of individual elastic constants, $C_{ij}$, are rare. In many cases, polycrystalline materials are studied experimentally for which one may determine the polycrystalline bulk modulus $(B)$, Young's modulus $(E)$ and shear modulus $(G)$. Using DLM-MD theory, we can calculate the elastic properties of a single crystal but by using Voigt and Reuss approaches, we can obtain expressions for bulk and shear moduli in polycrystals. For a cubic system these properties are derived as
\begin{equation}
\label{eq.8}
B_V=B_R=B
\end{equation}
\begin{equation}
\label{eq.9}
B=\frac{C_{11}+2C_{12}}{3}
\end{equation} 
\begin{equation}
\label{eq.10}
G_V=\frac{C_{11}-C_{12}+3C_{44}}{5}
\end{equation}     
\begin{equation}
\label{eq.11}
G_R=\frac{5(C_{11}-C_{12})C_{44}}{3(C_{11}-C_{12})+4C_{44}}
\end{equation}
The Young modulus $(E)$ and the Poisson ratio $(\nu)$ can also be calculated according to the following relations.
\begin{equation}
\label{eq.12}
E_{V,R}=\frac{9BG_{V,R}}{3B+G_{V,R}}
\end{equation}
\begin{equation}
\label{eq.13}
\nu_{V,R}=\frac{3B-2G_{V,R}}{2(3B+G_{V,R})}
\end{equation}
It is also useful to define the elastic anisotropy for polycrystalline materials.
\begin{equation}
\label{eq.14}
A_{V,R}=\frac{G_V-G_R}{G_V+G_R}
\end{equation}
The Voingt and the Reuss averaging of elastic constants, Eq.~\ref{eq.8}-\ref{eq.14}, will give us the upper and the lower bound of elastic constants, respectively. the Voigt-Reuss-Hill approach (the Hill approximation) combines these two limits by averaging over the Voigt and the Reuss elastic constants, assuming that this average gives a good approximation for the actual macroscopic elastic constants\cite{Hirsekorn1990,Toonder2000}.
\begin{equation}
\label{eq.h1}
E_{H}=\frac{E_{V}+E_{R}}{2}
\end{equation}
\begin{equation}
\label{eq.h2}
G_{H}=\frac{G_{V}+G_{R}}{2}
\end{equation}
and
\begin{equation}
\label{eq.h3}
\nu_{H}=\frac{E_{H}}{2G_{H}}-1
\end{equation}


\subsection{Details of DLM-MD Simulations} \label{2B}
\begin{figure}[!t]
  \centering
    \includegraphics[width=1.0\textwidth]{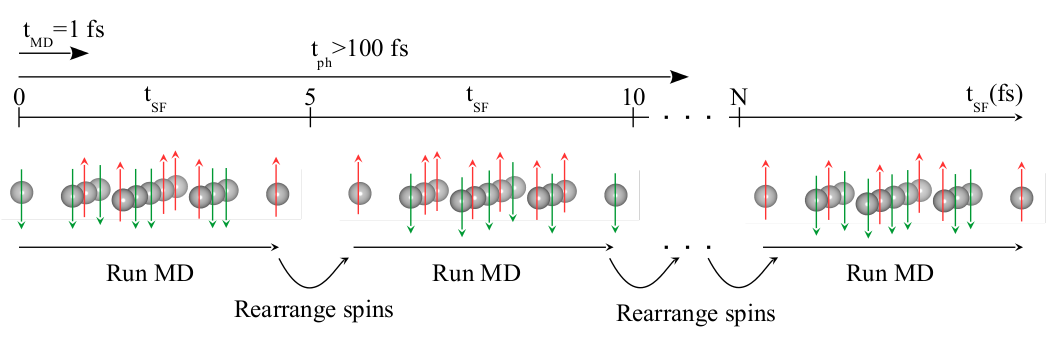}
    \caption{(Color online) Schematic representation of DLM-MD process. Starting with a random arrangements of magnetic moments, an MD calculation is run for $N_{MD}^{SF}=t_{SF}/t_{MD}$ steps during which the magnetic configuration is allowed to evolve according to the solution of the electronic structure problem. At the $t_{SF}$, the moments are flipped and rearranged in another random magnetic configuration, while positions and velocities of all the atoms are preserved. Then the MD simulation continues. The figure shows only one layer of Cr atoms from the CrN supercell and the N atoms are not shown here. The spin-flip time is chosen to be $t_{SF}=5$ fs. The Born-Oppenheimer MD time step is chosen as $t_{MD}=1$ fs. The phonon time scale is also shown as $t_{ph} > 100$ fs. The lengths of the time arrows demonstrate that the time scale of the collective atomic vibrational modes are much slower than the timescales of both the electronic and transverse local magnetic moments ($t_{MD} \leq t_{SF} < t_{ph} $).}
\label{fig0}
\end{figure}

The DLM-MD method was suggested by Steneteg \textit{et al}. to simulate the paramagnetic state of magnetic materials at finite temperatures including lattice vibrations\cite{Steneteg2012a}. In this method, the disordered local moments (DLM) picture of paramagnetism is implemented in the framework of the \textit{ab initio} molecular dynamics (MD). The DLM approach is introduced by Hubbard\cite{Hubbard1981b,Hubbard1979d,Hubbard1979e} and Hasegawa\cite{Hasegawa1979b,Hasegawa1980b} and later on applied by Gyorffy \textit{et al.} within the coherent potential approximation (CPA) electronic structure framework\cite{Gyorffy1985b}. 
Beside the LDA+U approximation for the exchange correlation energy and the supercell description of the disordered magnetic state, additional approximations in DLM-MD approach need to be noted. First, we consider the local moments to have collinear orientations. Indeed it is well justified to ignore the noncollinear orientations of local moments in paramagnetic materials in a description of thermodynamic potentials, well above the magnetic transition temperature\cite{Gyorffy1985b}. The fast longitudinal fluctuations of magnetic moments dictating their magnitude are treated as fast electronic degrees of freedom governed by the self-consistent solution of the electronic structure problem at each MD step. \\
Following the same arguments as in Ref.~\citen{Gyorffy1985b}, we assume that the magnetic configuration of the system, even though ergodic, does not cover its phase space uniformly in time but rather gets stuck for a specified time (we denote as spin-flip time, $t_{SF}$) near the points with a specific configuration of moments at every site pointing in a random direction and then moves instantly to another disordered but different state of the phase space. In fact, Gyorffy \textit{et al.} supposed that the changes in the orientational configuration of the moments characterize the motion of temporarily broken ergodicity to a large degree.\\
It should be noted that on one hand, in the paramagnetic state at high temperatures the magnetic excitations exhibit different characteristics from those of low temperature. As an important example, the relevant time scale of its dynamics can be better estimated by the spin decoherence time, $t_{dc}$, of individual or pair of moments, rather than the inverse spin-wave frequency related to the collective motion of many spins. This means that the relevant magnetic dynamics are much faster in magnetically disordered materials (of the order of $10^{-14}$-$10^{-15}$ s) in comparison to that of in magnetically ordered systems (of the order of $10^{-13}$ s). \\
A further complication is that the paramagnetic state is a high temperature state of a magnetic material for which considering the atomic motions and displcements from ideal lattice sites in the simulations is essential\cite{Alling2016}. The time scale for collective atomic motions constituting lattice vibrations is estimated by the inverse of the Debye frequency ($\sim 10^{-12}$). However, in molecular dynamics simulations, we need to calculate the displacements and forces acting on individual atoms on a much shorter time scale of the order of $10^{-15}$ s for a proper description of Born-Oppenheimer (B-O) dynamics.\\
Figure.~\ref{fig0} illustrates schematically the basic idea of the DLM-MD technique and demonstrates all the different timescales of the method including the B-O MD time step ($t_{MD}$), the spin-flip time ($t_{SF}$) as well as a time indicator for the "phonon timescales" corresponding to the period time of the highest frequency phonons. The values we have adopted in our simulations, as shown in the figure, comply with the Born-Oppenheimer approximation indicating that electrons exhibit the fastest timescale as compared to other degrees of freedom. Then comes the timescale of the individual atomic motions ($t_{MD}=1$ fs). The change in the temporarily broken ergodicity magnetic state of the disordered local magnetic moments comes next ($t_{SF}=5$ fs) and only then comes the timescale of the full collective atomic vibrational modes ($t_{ph}>100$ fs).\\
Strictly speaking, the appropriate value of the spin flip time should either be taken from experiments or calculated from real spin dynamics simulations. However, in Ref.~\citen{Steneteg2012a} it was shown that $t_SF$ should be chosen short enough to assure an adiabatic approximation. In particular, several tests with different spin flip times carried out in Ref.~\citen{Steneteg2012a} demonstrate that $t_SF=5$ fs gives reliable description for CrN.  

We run DLM-MD simulations with total number of 5000 steps, resulting in a length of 5 ps for one simulation. The schematic representation of this algorithm is shown is Fig.~\ref{fig0}. \\
Using DLM-MD, the elastic constants for PM phase is calculated at five different temperatures 300, 600, 800, 1000, 1200. For each temperature, five different values of distortions, $\eta \in \{-0.02, -0.01, 0.00, 0.01, 0.02\}$ have been used for the deformation matrix $\epsilon(\eta)+I$. We can then extract the $\sigma$ values, using Eq.~\ref{eq.5}-\ref{eq.7} and calculate the elastic constants at each temperature.\\
We note that for DLM-MD calculations of $C_{ij}$ constants, we do not need to use the projected cubic elastic constants as described in Sec.~\ref{sec.2C}, as we sample the phase space of possible magnetic configurations and average out non-cubic magnetic symmetry on the fly.  


\subsection{Finite Temperature Elastic Constants From Static Calculations Including Thermal Expansion Effects}
\label{sec.2C}
In order to simulate the paramagnetic state of CrN in a static lattice approximation at T=0 K, we have used the DLM picture combined with magnetic special quasirandom structure (SQS) approach\cite{Alling2010d}. In our static DLM-SQS calculations, we have employed a $3\times3\times3$ cubic supercell with 108 Cr and 108 N atoms in which the spin-up and spin-down Cr moments are mimicing a random alloy distribution\cite{Zunger1990}. \\
To obtain the elastic properties, we apply a set of different distortions to this supercell. As explained in Sec.~\ref{2A}, after performing the first-principles calculations for each of these supercells, we calculate the stress. Thereafter, we obtain the derivative of the stress numerically. This derivative will provide us with different $C_{ij}$ values. \\
However, one should bear in mind that due to magnetic disorder the cubic symmetry of the supercell is broken and all the elements of the elastic matrix, $C_{11}$, $C_{22}$, $C_{33}$, $C_{12}$, $C_{13}$, $C_{23}$, $C_{44}$, $C_{55}$ and $C_{66}$ are not identical and need to be calculated. Then we use the projection technique to determine average cubic elastic constants\cite{Tasnadi2012} of the simulated cubic crystal via
\begin{equation}
\bar{C_{11}}=\frac{C_{11}+C_{22}+C_{33}}{3}\label{eq.av1}
\end{equation} 
   \begin{equation}
\bar{C_{12}}=\frac{C_{12}+C_{13}+C_{23}}{3}\label{eq.av2}
\end{equation}
\begin{equation}
\bar{C_{44}}=\frac{C_{44}+C_{55}+C_{66}}{3}\label{eq.av3}
\end{equation}

Thermal expansion is one important manifestation of anharmonicity. Statically, one can include the temperature effect on the elastic properties through the thermal expansion\cite{Shang2010,Wang2010}. This approach is based on Birch's law\cite{Birch1961} that includes temperature effects in elastic properties implicitly via thermal expansion\cite{Ledbetter2006}. Thus within this approach
\begin{equation}
C_{ij}(T)\approx C^{0}_{ij}(V(T))\label{eq.15}
\end{equation}
where $C_{ij}^{0}(V(T))$, are the elastic constants calculated at zero temperature but at volume $V$ corresponding to simulation temperature $T$. Note that, for these set of calculations, we should use the average elastic constants as stated in Eq.~\ref{eq.av1}-\ref{eq.av3} since we are considering a magnetically disordered state. In this work, we have used the experimental thermal expansion obtained from Ref.~\citen{Wang2012a}.


\subsection{Finite Temperature Elastic Constants From SIFC-TDEP}
We concentrate on mechanical properties calculated for single crystal materials from first principles simulations. The elastic constants can be defined either by Hooke's law, described above or in long wave limit of phonons. In this section we give a brief description of the calculation of the elastic properties through the phonons and introducing the methods that include explicit temperature dependence of phonon properties and therefore elastic constants. 

Let us Start with a harmonic Hamiltonian describing the lattice dynamics
\begin{equation}\label{eq:model_ham}
\begin{split}
\hat{H}= & U_0+\sum_{i\alpha}\frac{p^2_{i\alpha}}{2 m_i}+
\frac{1}{2!}\sum_{ij}\sum_{\alpha\beta}
\Phi_{\alpha\beta}^{ij}
u_{i\alpha} u_{j\beta}\,,
\end{split}
\end{equation}
where $\alpha$ and $\beta$ are indices for the Cartesian coordinates and $u_\alpha$ and $u_\beta$ are the Cartesian components of the displacement of the atom $i$ and $j$, and $U_0$ is the potential energy of the static lattice; $p$ and $m$ are the momentum and mass of atom $i$. The $\Phi_{\alpha\beta}^{ij}$ are the interatomic force constants, which express a relation between the force in $\alpha$ direction acting on the atom $i$, when atom $j$ is displaced in direction $\beta$. 

In long wave limit $q \longrightarrow 0 $, the atoms move very slowly with vanishing frequencies and these sound waves have frequencies that are determined by macroscopic elastic constants. We can write the relation between elastic and force constants\cite{Leibfried1961,Fultz2010,Maradudin1962}
\begin{equation}\label{eq:d_fc}
  \tilde{C}_{ijkl}=-\frac{1}{2V}
\sum_{n} \Phi^{ij}_{0n} r^k_n r^l_n,
\end{equation}
where $\tilde{C}_{ijkl}$ is a full elastic constants matrix, that could be later reduced due to the symmetry and transformed to the Voigt notation. Here $\Phi^{ij}_{0n}$ the interatomic force constants (IFC) between atoms $i$ and $j$ in the unit cell $n$; $r^k_n$ and $r^l_n$ are the positions of atoms $k$ and $l$ in the unit cell n taken in respect to the reference unit cell 0.
To calculate interatomic force constants we use the temperature dependent effective potential (TDEP) method \cite{Hellman2013,Hellman2011}. The advantage of this method is that the temperature dependence of the IFC is included explicitly. We determine the force constants by minimising the difference in forces from the model Hamiltonian (Eq. \eqref{eq:model_ham}) and real system simulated with the DLM-MD. The DLM-MD provides us with a set of displacements $\set{\vec{U}_t^{\textrm{DLM-MD}}}$ and forces $\set{\vec{F}_t^{\textrm{DLM-MD}}}$ at each step, so we minimize the following quantity 
\begin{equation}\label{eq:min_f_phi}
\begin{split}
\min_{\mat{\Phi}}\Delta \vec{F} & =\frac{1}{N_t} \sum_{t=1}^{N_t}  \left| \vec{F}_t^{\textrm{DLM-MD}}-\vec{F}_t^{\textrm{H}} \right|^2= \\
& =\frac{1}{N_t} \sum_{t=1}^{N_t} \left| \vec{F}_t^{\textrm{DLM-MD}}-\mat{\Phi}(\vec{U}_t^{\textrm{DLM-MD}}) \right|^2,
\end{split}
\end{equation}
The solution for the force constants $\mat{\Phi}$ is obtained by linear square method that minimises $\Delta \vec{F}$. The symmetry constraints on the force constants matrices are applied \cite{Hellman2013,Hellman2011} and this procedure reduces the computational cost. Using TDEP one can get temperature$/$volume dependent IFC. TDEP works for the ordered structures. In disordered systems the symmetry of the crystal is broken. Therefore, we use a generalization  of the TDEP method towards disordered systems, the so-called SIFC-TDEP\cite{Shulumba2015}. Here it is applied to calculate vibrational properties at finite temperatures for magnetically disordered systems and to evaluate their elastic properties. Using the SIFC-TDEP method we extract the effective interatomic force constants ($\mat{\Phi}^\mathrm{{eff}}$) by treating Cr atoms as symmetry equivalent and imposing the full symmetry of the underlying crystal lattice on the IFCs. Obtained effective IFCs are then used to calculate elastic constants (Eq. \eqref{eq:d_fc}).

It is well known that there is an issue of using Eq. \eqref{eq:d_fc} to calculate accurately absolute values of the finite temperature elastic constants, because the result depends on the finite size of the simulation cells. On the other hand, it has been demonstrated that using the SIFC-TDEP and the original TDEP we are able to accurately determine the temperature dependence of phonon frequencies of CrN\cite{Shulumba2014} and other materials\cite{Li2014,Romero2015}. 

Therefore, we employ the finite temperature scaling of elastic constants to obtain the final expression, given by  
\begin{equation}
C_{ij}(V,T)=C_{ij}^{stat}(V,T_0)\frac{C_{ij}^{ph}(V,T)}{C_{ij}^{ph}(V,T_0)} \label{eq.16}
\end{equation}       
with $T_0=300$ K in our calculations. Note that one can avoid the scaling of the elastic constants by performing calculations on a larger cell. Thus elastic constants from SIFC-TDEP can be calculated including both volume and temperature dependence of IFC, extracted from SIFC-TDEP.

SIFC-TDEP scheme allows one to calculate the finite temperature elastic constants at a fraction (1/5) of computational cost of DLM-MD, in which the supercell has to be distorted using deformation matrix, Eq.~\ref{eq.2}. Shulumba \textit{et al.}\cite{Shulumba2015} showed that $\sim$80\% of the temperature effect on the elastic constant of TiN could be captured with this method, at 20\% of the computational cost. In this work, we extend the approach towards the magnetically disordered systems. 


\subsection{Computational Details}
All the static DLM and DLM-MD calculations are carried out within the projected augmented wave method (PAW)\cite{Blochl1994} as implemented in Vienna Ab-initio Simulation Package (VASP)\cite{Kresse1993a,Kresse1996a,Kresse1996b,Kresse1999a}. For the electronic exchange-correlation effects we have used a combination of local density approximation with a Hubbard Coloumb term (LDA+U)\cite{Anisimov1991a}. The effective Hubbard term value, $U^{eff}=U-J$, is chosen to be 3 eV for Cr $3d$ orbitals which is shown to be the optimal value obtained from a thorough theoretical comparison of the structural and electronic properties of CrN with experimental measurements\cite{Alling2010d}. The energy cutoff is set to 500 eV. \\  
We have used a supercell consisting of $3\times3\times3$ repetitions of the conventional cubic cell including 8 atoms, giving in total 108 Cr and 108 N atoms. In the PM phase, the spin-up and spin-down magnetic moments are randomly distributed on Cr atoms. The Brillouin zone is sampled using a Monkhorst-Pack scheme\cite{Pack1977} with a $k$-mesh of $3\times3\times3$. In order to maintain the desired temperature in our MD calculations, we have used the canonical ensemble (NVT). We have used the Nose thermostat\cite{Nose1991a} with the default mass value as it is implemented in VASP in our simulations.\\
The thermal expansion is included in our calculations using the experimental lattice constants as a function of temperature\cite{Wang2012a}. 

For SIFC-TDEP calculations we run DLM-MD with the same setting as described above with a difference in supercell size. We used a supercell contained 32 chromium and 32 nitrogen atoms arranged in 2$\times$2$\times$2 conventional unit cells. We ran the simulations on a grid of 6 temperatures and six volumes the NVT ensemble. From these simulations we extract IFC as a function of volume and temperature. The effective IFC were found to be smooth and easily interpolated across the whole temperature/volume interval. We calculate the theoretical thermal expansion, which is in good agreement with the experimental values\cite{Wang2012a}. Elastic constants from SIFC-TDEP are evaluated for the theoretical thermal expansion.\\
When dealing with MD simulations, we should make sure that the obtained results are well converged, i.e. that the statistical errors are small. The output of an MD run is reported in terms of of the time average, in our case of the elastic constants. Since the simulation times are of finite size, an statistical imprecision of this average values is expected. Our simulation time is 5 ps and in order to estimate the uncertainty of our MD results, we have used a \textit{t} distribution with 95\% confidence interval taking the correlated nature of each MD time step into account according to the method suggested by Allen and Tildesley\cite{Allen1991}. In our case we find that the factor of uncorrelated time steps correspond to between 1 to 200 configurations, which add completely new information to the average values, depending on the temperature. The 95\% confidence interval for the mean values of elastic constants and derived properties are given as error bars calculated in this way. 
\begin{table*}[t]
\caption{Temperature dependent elastic constants of PM B1 CrN obtained by different methods including experimental, static (T=0 K) nonmagnetic (NM) and antiferromagnetic (AFM) values for comparison.}
\begin{center}
\begin{tabular}{l|cccccccccccc}
\hline\hline
\diagbox{Method}{Elast. Const.(GPa)} && $C_{11}$ && $C_{12}$ && $C_{44}$ && $B_{V}=B_{R}$ && $E_{V}$/$E_{R}$ && $G_{V}$/$G_{R}$\\
\hline
Static, PM B1 (T=0 K)\cite{Zhou2014} && 649 && 99 && 145 && 282 && 479/443 && 197/179 \\
\hline
Static, PM B1 (T=0 K, This work) && 624 && 98 && 141 && 273 && 462/428 && 189/173\\
\hline
Static, AFM B1 (T=0 K, This work) && 659 && 108 && 144 && 292 && 481/443 && 196/178\\
\hline
Static, NM B1 (T=0 K)\cite{Zhou2014} && 641 && 260 && -59 && 387 && 118/-416 && 41/-124\\
\hline
Experiment (T=300 K) && 540\cite{Almer2003,Chen2004} && 27\cite{Almer2003} && 88\cite{Almer2003} &&  && 400 \cite{Holleck1986} && \\
\hline
DLM-MD (T=300 K) && 591 && 102 && 141 && 265 && 446/420 && 183/170 \\
\hline
DLM-MD (T=1200 K)  && 486 && 108 && 132 && 234 && 381/371 && 155/150\\
\hline
SIFC-TDEP (T=1200 K) && 516 && 95 && 135 && 235 && 403/388 && 166/158\\
\hline
\end{tabular}
\end{center}
\label{tab.1}
\end{table*}


\section{Results}
\subsection{Single crystal Elastic Constants}
\begin{figure}[!ht]
  \centering
    \includegraphics[width=0.8\textwidth]{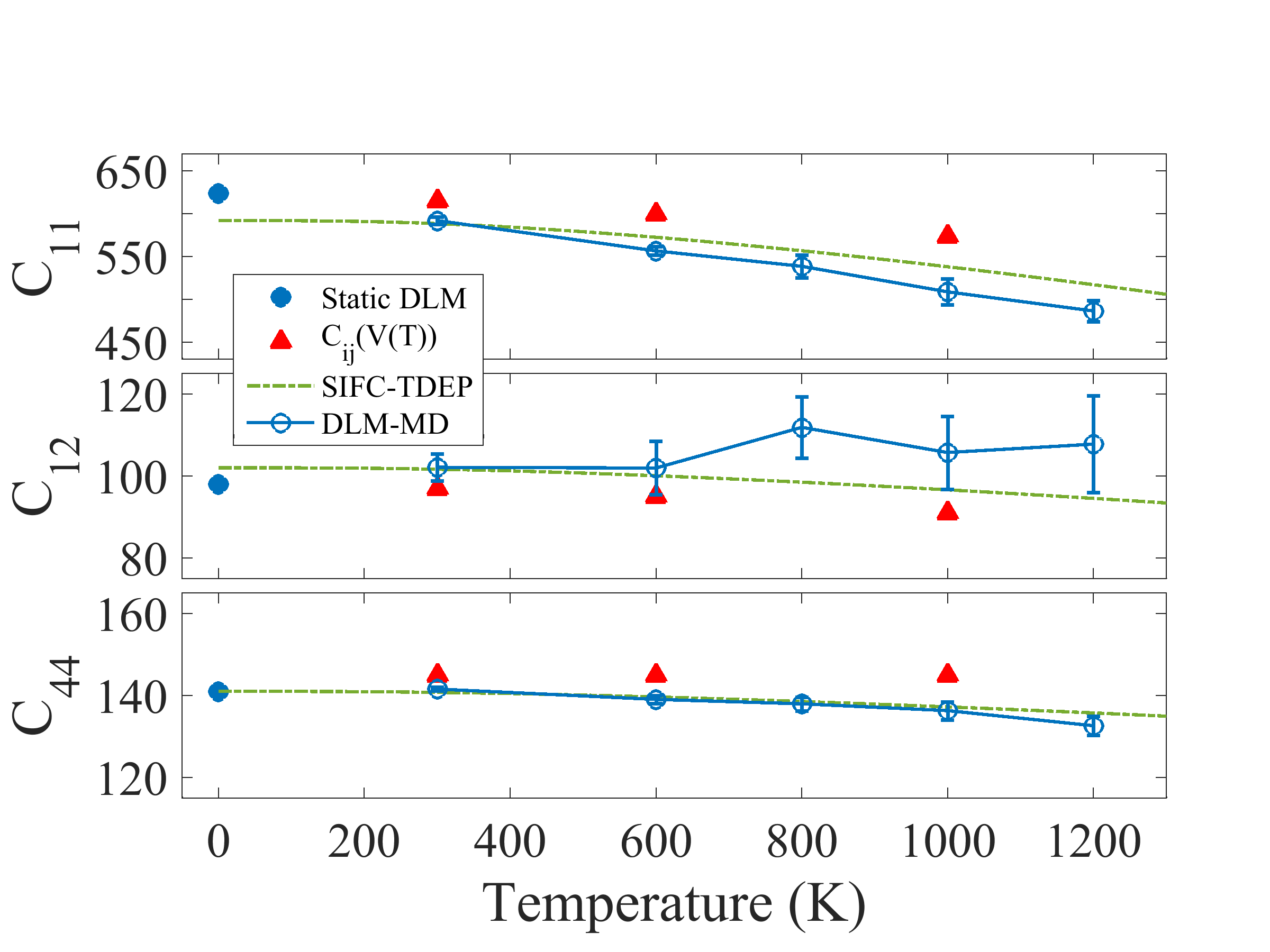}
    \caption{(Color online) Calculated temperature dependent single-crystal elastic constants of PM CrN obtained by using different theoretical methods. The errorbars correspond to the standard deviation of the molecular dynamics simulations with 95\% confidence interval.}
\label{fig1}
\end{figure}

Figure~\ref{fig1}, shows the obtained temperature dependent elastic constants of the paramagnetic B1 CrN from different methods. We can see that the low-temperature limit of the elastic constants calculated at finite temperatures via DLM-MD method are in good agreement with the corresponding elastic constants obtained from static zero-temperature calculations. The zero-Kelvin values also agree well with earlier theoretical calculations (see, for instance, Ref.\cite{Zhou2014}). 

In Tab.~\ref{tab.1} we report the calculated elastic properties of nonmagnetic (NM), antiferromagnetic (AFM) and paramagnetic (PM) B1-CrN at zero kelvin. The values of the elastic constants from our AFM and PM calculations show that AFM calculations in general tend to overestimate the elastic constants. Relatively small difference between AFM and PM calculations can be attributed to relatively weak magnetic exchange interactions in B1 CrN\cite{Lindmaa2013a}. One should expect substantially larger effect in systems with higher Ne\'{e}l or Curie temperature. Simultaneously, from the values listed in Tab.~\ref{tab.1} we can see that the nonmagnetic calculations give a negative value for $C_{44}$. This implies that NM B1-CrN is mechanically unstable as the $C_{ij}$ values does not satisfy the Born-Huang stability criterion\cite{Born1988} for which the shear elastic coefficient $C_{44}$ should be positive. Most other elastic moduli come out very wrong as well in non-magnetic calculations. This indicates the importance of magnetic effects which need to be considered while simulating the PM state of a magnetic material. \\   
Tab.~\ref{tab.1} also summarizes the elastic properties of PM B1 CrN at T=300 K. We observe that our calculations overestimate $C_{11}$ by 51 GPa, $\sim$9\%, as compared to the experimental value\cite{Almer2003,Chen2004} which is reasonable given LDA+U normal uncertainty. The difference between the theoretical and experimental values for $C_{44}$ is 53 GPa, $\sim$38\%, and 75 GPa, $\sim$74\% for $C_{12}$. However, we do not trust the experimental values of $C_{12}$ and $C_{44}$ and emphasize that our results are in good agreement with other first-principles calculations\cite{Zhou2014}. Moreover, our average Young's modulus value at 300 K, E=433 GPa is in good agreement with the reported experimental value of 400 GPa\cite{Holleck1986}. Considering this excellent agreement between theory and experiment for E, additional independent measurements of the $C_{ij}$ values at least at room temperature are desired. \\
Note that the numerical accuracy of elastic constants calculated with DLM-MD method is quite high. The statistical errors for $C_{11}$ and $C_{44}$ are at most within 0.1\% of the mean values which is very small. Both $C_{11}$ and $C_{44}$ decrease almost linearly with increasing temperature. This indicates the normal temperature dependence behavior originating from anharmonicity\cite{Grimvall1999}. For $C_{12}$, on the other hand we do not see any specific trend and it appears to be nearly temperature independent within the error bars. SIFC-TDEP values with all the elastic constants decrease monotonously. \\
For many systems in the absence of phase transitions for intermediate temperatures, the temperature dependent elastic constants can be fitted to the empirical relation\cite{Grimvall1999} 
\begin{equation}
C_{ij}(T)=C_{ij}(0)(1-b(T-T_{N})) \label{eq.17}
\end{equation}
where $b$ is a constant. As we get close to the melting temperature, high-order anharmonic effects result in a strong nonlinear temperature dependence\cite{Grimvall1999}. As for the case of CrN, the $T_{Ne\'el}\sim 286$ $K$, we have $T_N < T < T_m$, in which T is the temperature range in which we do our simulations and $T_m\sim 1500$ K is the melting temperature of CrN. This argument further justifies the reliability of our method because our simulations result in a nearly linear (within numerical accuracy) temperature dependence of elastic constants.\\
The finite temperature values from DLM-MD are in good agreement with the data obtained from SIFC-TDEP.
\begin{figure}[!ht]
  \centering
    \includegraphics[width=0.8\textwidth]{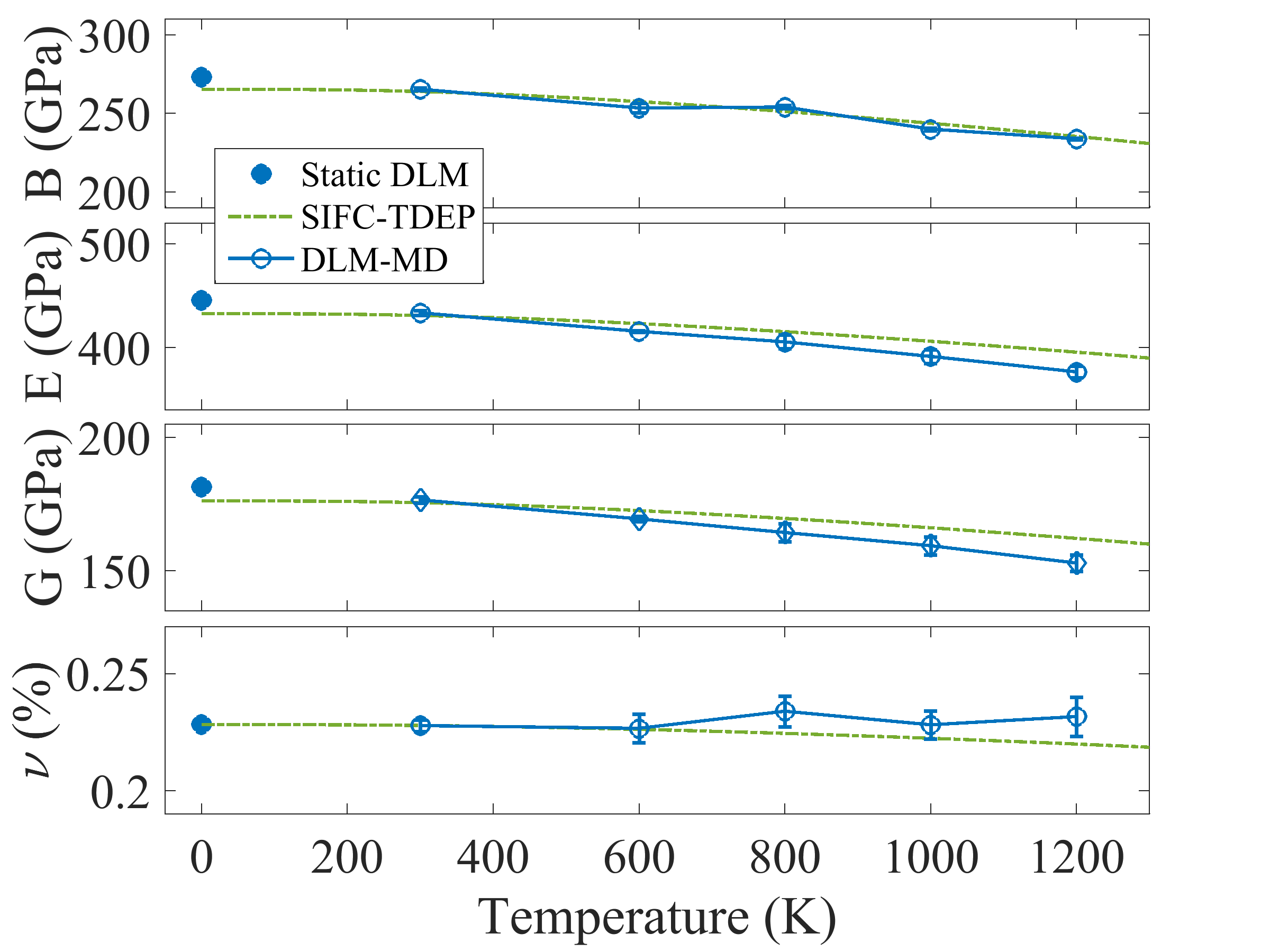}
    \caption{(Color online) Calculated Voigt-Reuss-Hill averages (Eq.~\ref{eq.h1}-\ref{eq.h3}) of Polycrystalline elastic constants of PM CrN from top to bottom (a) Bulk modulus, (b) Young's Modulus, (c) Shear Modulus and (d) Poisson ratio, as a function of temperature. The errorbars correspond to the standard deviation of the molecular dynamics simulations with 95\% confidence interval.}
\label{fig2}
\end{figure}

The single crystal elastic constants of PM CrN at 1200 K is given in Table.~\ref{tab.1}. As can be seen from Fig.~\ref{fig1}, DLM-MD calculations at T=300 K give 591 GPa, 102 GPa and 141 GPa for $C_{11}$, $C_{12}$ and $C_{44}$, respectively. At T=1200 K, Table.~\ref{tab.1}, SIFC-TDEP gives a larger value of $C_{11}=516$ GPa which differs by 23 GPa, $\sim$6\% from $C_{11}=486$ GPa from DLM-MD. The $C_{12}=95$ GPa is smaller by 15 GPa, $\sim$12\% in SIFC-TDEP as compared to $C_{12}=108$ GPa in DLM-MD. The $C_{44}$ is 132 GPa and 135 GPa from DLM-MD and SIFC-TDEP, respectively. Clearly there is an increasing difference between DLM-MD and SIFC-TDEP values as the temperature increases. On the other hand, SIFC-TDEP clearly shows much higher numerical stability and much smoother temperature dependence as compared to the DLM-MD calculations, which shows the usefulness of this numerically efficient technique for calculations of the temperature elastic constants in a not too broad temperature interval. \\
The red triangles in Fig.~\ref{fig1}, show the results from DLM static calculations including thermal expansion. The method gives close values of $C_{12}$ in comparison to DLM-MD and SIFC-TDEP $C_{12}$, but the data for other elastic constants, $C_{11}$ and $C_{44}$ differ stronger. This divergence demonstrates that the Birch law may be violated in real systems at finite temperatures. 
This implies that incorporating the temperature effect through thermal expansion may not be an accurate way to obtain the temperature dependence of elastic properties of magnetic materials in their paramagnetic state.
In order to get a good picture for finite temperature elastic properties in PM CrN, we need to treat lattice vibrations, magnetic configuration and the effect of thermal expansion on the same footing as it is done in our DLM-MD simulations.


\subsection{Polycrystalline Elastic Constants}
Using single crystal elastic constants, obtained from our methods, we can calculate the temperature dependent polycrystalline elastic constants and the Poisson ratio for PM B1 CrN. The results are displayed in Fig.~\ref{fig2}. Similar to what we see in Fig.~\ref{fig1} for $C_{ij}$, B, E and G moduli show nearly linear temperature dependence following eq.~\ref{eq.17}. Poisson ratio shows a little bit of variation as the temperature increases. As stated for the single crystal elastic constants, we see a good agreement between the room temperature values obtained from our DLM-MD calculations and from conventional zero kelvin static calculations. The polycrystalline elastic constants values obtained from DLM-MD and SIFC-TDEP are close to each other suggesting that both methods give similar results. \\
At zero kelvin, the bulk modulus (B) from AFM calculations, as listed in Tab.~\ref{tab.1}, is found to be 292 GPa which is significantly larger than the PM value of 273 GPa. There is even a larger difference between the values of bulk moduli obtained from PM and NM calculations. The nonmagnetic calculations give the bulk modulus of 387 GPa. However, it is shown in other studies that simulating the paramagnetic state as nonmagnetic could lead to erroneous conclusions. Rivadulla \textit{et al.} carried out calculations on CrN considering the PM state as nonmagnetic. Their study gave a bulk modulus value of 340 GPa for the cubic CrN showing a drastic reduction of about 25\% as compared to that of the orthorhombic phase (255 GPa)\cite{Rivadulla2009a}. Alling \textit{et al.}\cite{Alling2010e} calculated the bulk modulus of cubic CrN using DLM to simulate the paramagnetic state and they found the bulk modulus to be 252 GPa which is very close to the bulk modulus of the orthorhombic phase. These results highlight the importance of taking into account the finite local moments during the simulation of the paramagnetic state.\\
There are several experimental studies on the Young's modulus of CrN measured for thin films \cite{Holleck1986,Attar1995} with preferred orientations\cite{Attar1995,Cunha1999,Chen2004}. The reported experimental values for CrN range from 324 GPa to 461 GPa. Our obtained values from room temperature calculations are summarized in Tab.~\ref{tab.1}. We observe that our results from both DLM-MD and SIFC-TDEP, are well within the range of reported experimental values. Our average static Poisson ratio value, $\nu\sim0.25$, is in good agreement with the values derived from experimental data, 0.28\cite{Fabis1990,Cunha1999a,Qiu2013} and 0.24\cite{Chen2004}. From Fig.~\ref{fig2}, we can observe that all polycrystalline elastic constants have a fairly strong temperature dependence, decreasing by almost $\sim$14\% in DLM-MD and $\sim$8\% in SIFC-TDEP between room temperature and 1200 $K$. 
\begin{figure}[!ht]
  \centering
    \includegraphics[width=0.8\textwidth]{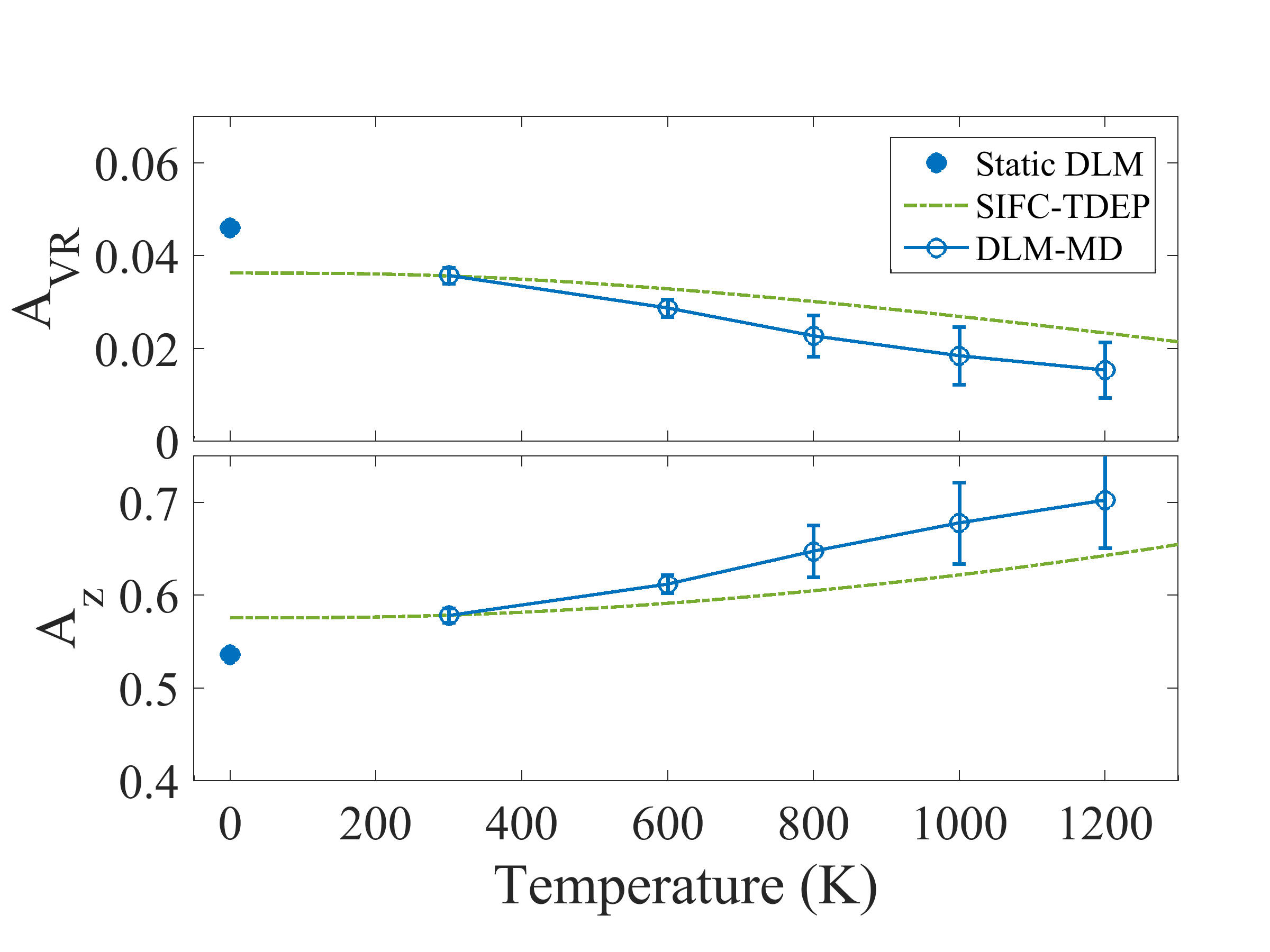}
    \caption{(Color online) Calculated Voigt-Reuss-Hill Anisotropy, $A_{VR}$ and Zenner elastic shear, $A_Z$, of PM CrN as a function of temperature. The errorbars correspond to the standard deviation of the molecular dynamics simulations with 95\% confidence interval.}
\label{fig3}
\end{figure}


\subsection{Elastic Anisotropy} 
It is possible to determine the measure of the elastic anisotropy experimentally by the strain ratio. The strain ratio can also be recalculated by the ratio between the Young's moduli $E_{hkl}$ in different directions\cite{Tasnadi2010}. We can calculate the $E_{hkl}$ from the single-crystal elastic constants\cite{Grimvall1999}. Our DLM-MD simulations at room temperature give $E_{111}=347$ GPa, $E_{200}\sim556$ GPa and $E_{220}\sim385$ GPa. Even though, there is a wide discrepancy in the experimentally measured directional Young's modulus values\cite{Chen2004}, our data are fairly consistent with the experimental values 290 GPa\cite{Cunha1999}, 520 GPa\cite{Attar1995} and 300$\pm$20 GPa\cite{Chen2004} for $<111>$, $<200>$ and $<220>$ directions, respectively. \\
In order to quantify the temperature dependence of elastic anisotropy, we calculated the anisotropy according to Voigt-Reuss-Hill definition, Eq.~\ref{eq.14}, as well as according to Zener\cite{Grimvall1999}, Fig.~\ref{fig3}.
\begin{equation}
A_Z=\frac{2C_{44}}{C_{11}-C_{12}}
\end{equation}  
If the material is isotropic, the former is equal to 0 as the temperature increases and the latter is equal to 1. In Fig.~\ref{fig3} we see that CrN becomes more isotropic at higher temperatures. 


\section{Summary and Conclusion}
We have used first-principles simulations based on \textit{ab initio} molecular dynamics (AIMD) in combination with disordered local moments (DLM) method to  study the finite temperature elastic properties of magnetic material CrN, in its high-T paramagnetic state. Though these simulations are computationally expensive and are substantially more time consuming as compared to conventional static calculations, we see that performing MD to obtain finite temperature elastic constants is needed to get a proper description of the elastic constants at elevated temperatures. Moreover, we have used the recently developed method, SIFC-TDEP, to study temperature dependent elastic constants of CrN. This method also uses DLM-MD but allows one to calculate elastic constants without simulations at distorted lattices. Thus it has higher computational efficiency. In general, we see that both DLM-MD and SIFC-TDEP give results that are in good agreement with each other and with available experiment. On the other hand, the use of Birch law may give larger errors for calculated elastic constants.\\
We study the temperature dependent elastic properties of prototypical paramagnetic transition metal nitride, CrN, between room temperature and 1200 $K$ which corresponds to operation temperature of cutting tools. We have calculated the single crystal elastic constants of PM cubic CrN, $C_{11}$, $C_{12}$ and $C_{44}$, as well as its polycrystalline elastic constants, B, G and E being the bulk, shear and Young's moduli, respectively and also the Poisson ratio, $\nu$. We observe that the elastic constants decrease nearly linearly with increasing temperature which is the predicted temperature dependent behavior, caused by anharmonicity. We see that polycrystalline elastic constants decrease by $\sim14\%$ between room temperature and 1200 $K$. Studying the elastic anisotropy, demonstrates that the material becomes substantially more isotropic at elevated temperatures.  Therefore, the effect of temperature on elastic properties is strong and should be included in the studies of materials functioning at high-T environments. The proposed technique allows for a reliable inclusion of finite temperature effects in \textit{ab initio} simulations of elastic properties of magnetic materials.

\section*{Acknowledgments}
E.M. would also like to asserts her appreciations to Dr. Ferenc Tasnadi for the very useful discussions. Financial support from the Swedish Research Council (VR) through Grant No. 621-2011-4426  and the Swedish Foundation for Strategic Research (SSF) program SRL Grant No. 10-0026 is Acknowledged. I.A.A. is grateful for the Grant from the Ministry of Education and Science of the Russian Federation (Grant No. 14.Y26.31.0005) and Tomsk State University Academic D. I. Mendeleev Fund Program. B. A. acknowledges financial support by the Swedish Research Council (VR) through the young researcher grant No. 621-2011-4417 and the international career grant No. 330-2014-6336 and Marie Sklodowska Curie Actions, Cofund, Project INCA 600398. Moreover, we acknowledge support from the Swedish Government Strategic Research Area in Materials Science on Functional Materials at Linko\"oping University (Faculty Grant SFOMatLiU No 2009 00971). The authors would like to acknowledge the Swedish National Infrastructure for Computing (SNIC) at the National Supercomputer Center (NSC) for providing the computational resources. 


\bibliography{/home/elin/Documents/library/Paper5-ref}
\end{document}